\documentclass[twocolumn,floatfix,prb,aps,showpacs,superscriptaddress]{revtex4-2}
\usepackage{graphicx,amsmath,amssymb,color,nicefrac,multirow, makecell, ragged2e,float,multirow,arydshln}
\usepackage[unicode=true,colorlinks=true,linkcolor=blue,citecolor=blue,urlcolor=blue]{hyperref}
\usepackage[titletoc,title]{appendix}

\newcommand{\be}{\begin{equation}}
\newcommand{\ee}{\end{equation}}

\newcommand{\ba}{\begin{eqnarray}}
\newcommand{\ea}{\end{eqnarray}}

\def\beq{\begin{eqnarray}}
\def\eeq{\end{eqnarray}}

\makeatletter
\newcommand*{\rom}[1]{\expandafter\@slowromancap\romannumeral #1@}
\makeatother

\newcommand{\non}{\nonumber\\}

\usepackage{bm,physics,enumerate,booktabs,ulem}

\newcommand{\tx}[1]{{\text{#1}}}

\usepackage{orcidlink}

\begin{document}
\title{
Topological superconductivity on a kagome magnet
coupled to a Rashba superconductor}

\author{Koji Kudo}
\affiliation{Department of Physics, Kyushu University, Fukuoka 819-0395, Japan}

\author{Ryota Nakai}
\affiliation{RIKEN Center for Quantum Computing (RQC), Wako, Saitama, 351-0198, Japan}

\author{Hiroki Isobe}
\affiliation{Department of Physics, Kyushu University, Fukuoka 819-0395, Japan}

\author{Kentaro Nomura}
\affiliation{Department of Physics, Kyushu University, Fukuoka 819-0395, Japan}
\affiliation{Quantum and Spacetime Research Institute, Kyushu University, Fukuoka 819-0395, Japan}

\begin{abstract}
 A quantum anomalous Hall system is predicted to realize 
 topological superconductivity when proximity-coupled to an $s$-wave superconductor.
 A kagome magnet with chiral magnetic ordering exhibits the quantum 
 anomalous Hall effect; however,
 superconducting 
 proximity to an ordinary $s$-wave superconductor fails to induce
 pairing
 in the strong exchange coupling limit.
 In this work, we demonstrate that proximity coupling to a
 Rashba superconductor gives rise to
 topological superconducting phases characterized by odd Bogoliubov-de Gennes 
 Chern numbers. 
 We confirmed their consistency with the chiral central charge calculated based on the modular commutator.
 We also show that the magnetic ordering of kagome magnets is affected 
 energetically by the proximity effect.
\end{abstract}

\maketitle

\section{Introduction}
Realizing Majorana zero modes, which are paradigmatic non-Abelian anyons, is
one of
the central goals of modern condensed matter physics. Several platforms have
been proposed, including the fractional quantum Hall 
effect~\cite{Tsui82,Laughlin83,Jain89,Moore91,Read99,Jain07}, unconventional 
superconductors~\cite{Read00,Ivanov01,Kitaev01}, and quantum spin 
liquids~\cite{Kalmeyer87,Levin05a,Kitaev06,Jackeli09}. 
A common requirement is the 
coexistence of nontrivial topology and superconducting pairing, whose interplay
produces topological superconductivity (TSC) with exotic quasiparticles. 
Rather than relying solely on intrinsically TSC,
another promising route is to engineer heterostructures in which a topological 
electron system, such as the surface of a topological 
insulator~\cite{Fu08,Akhmerov09}, a quantum Hall/quantum anomalous Hall 
system~\cite{Qi10,Bjorn16,Mishmash19,Jeon19,Chaudhary20,Schirmer22,Kudo24,Manna24,Nakai25,Kudo25,Antonenko25,Antonenko25b}
or magnetic systems~\cite{Choy11,Nadj-Perge13,Braunecker13,Klinovaja14,Vazifeh13,Pientka13,Nakosai13,Nadj-Perge14}, is proximity-coupled to an 
ordinary $s$-wave superconductor. The recent growing interest in 
two-dimensional van der Waals materials such as 
multilayer graphene~\cite{Sharpe19,Serlin20,Lu24,Kudo24b,Han25} 
and transition-metal dichalcogenides~\cite{Li21,Cai23,Zeng23,Park23,Xu23}, 
which exhibit integer and fractional quantum anomalous Hall insulators, has 
further stimulated the search for proximity-induced TSC.

Kagome materials are experimentally accessible systems that exhibit
rich electronic and magnetic properties  
~\cite{Mielke91,Mielke92,Sachdev92,Lecheminant97,Ohgushi00,Guo09,Balents10,Han12,Kudo17,Kudo19b,Kim19,Kobayashi19,Baidya20,Watanabe22,Tazai22,Yi23,Fujiwara23,Tazai23,Huang24,Tazai24,Shimura24,Asaba24,Nakazawa24,Nagashima25,Nakazawa25,Nakazawa25b,Onari25,Mojarro25,Sharma25,Wang25b,Yamakawa25,Tazai25,Tazai25b,Huang25b,Goto25,Hou25,Hou25b,Zhang26}. 
These properties arise from geometric frustration, Dirac dispersion, van Hove singularities, flat bands and so forth.
Kagome materials have been reported in a variety of systems, such as
Mn$_3$Sn~\cite{Nakatsuji15,Ito17,Liu17b,Kuroda17,Higo18,Higo18b,Park18},
Fe$_3$Sn$_2$~\cite{Ye18,Lin18,Fang22},
Mn$_3$Ge~\cite{Nayak16,Kiyohara16,Soh20,Jeon21,Jeon23},
Co$_3$Sn$_2$S$_2$~\cite{Liu18b,Wang18,Liu19b,Ozawa19,Ikeda21,Lau23,Nakazawa24b},
and CsV$_3$Sb$_5$~\cite{Ortiz19,Ortiz20,Zhao21c,Tan21,Nie22,Kang23,Wang23b,Murata25}.
Among these, kagome magnets -- kagome materials with magnetic orders such as 
Mn$_3$Ge -- have been shown to exhibit superconducting proximity effects when 
interfaced with Nb~\cite{Jeon21,Jeon23}.
These results highlight the feasibility of kagome magnets as a promising platform for proximity-induced 
TSC, since single-layer kagome magnets with chiral magnetic ordering have been suggested to exhibit
the quantum anomalous Hall effect~\cite{Ohgushi00}.

However, caution is required when itinerant magnetic materials with strong 
exchange coupling are proximity-coupled to conventional $s$-wave 
superconductors to induce TSC.
Indeed, single-layer kagome magnets are insusceptible to $s$-wave pairing in the limit of strong exchange coupling since
electrons at momenta $\bm{k}$ and $-\bm{k}$
are related by inversion symmetry, not by time-reversal symmetry, hence having parallel spins.
One 
way to realize TSC in such a system is to introduce a spatially nonuniform 
$s$-wave pairing~\cite{Chaou25} that breaks inversion symmetry. 

\begin{figure}[t]
\includegraphics[width=\columnwidth]{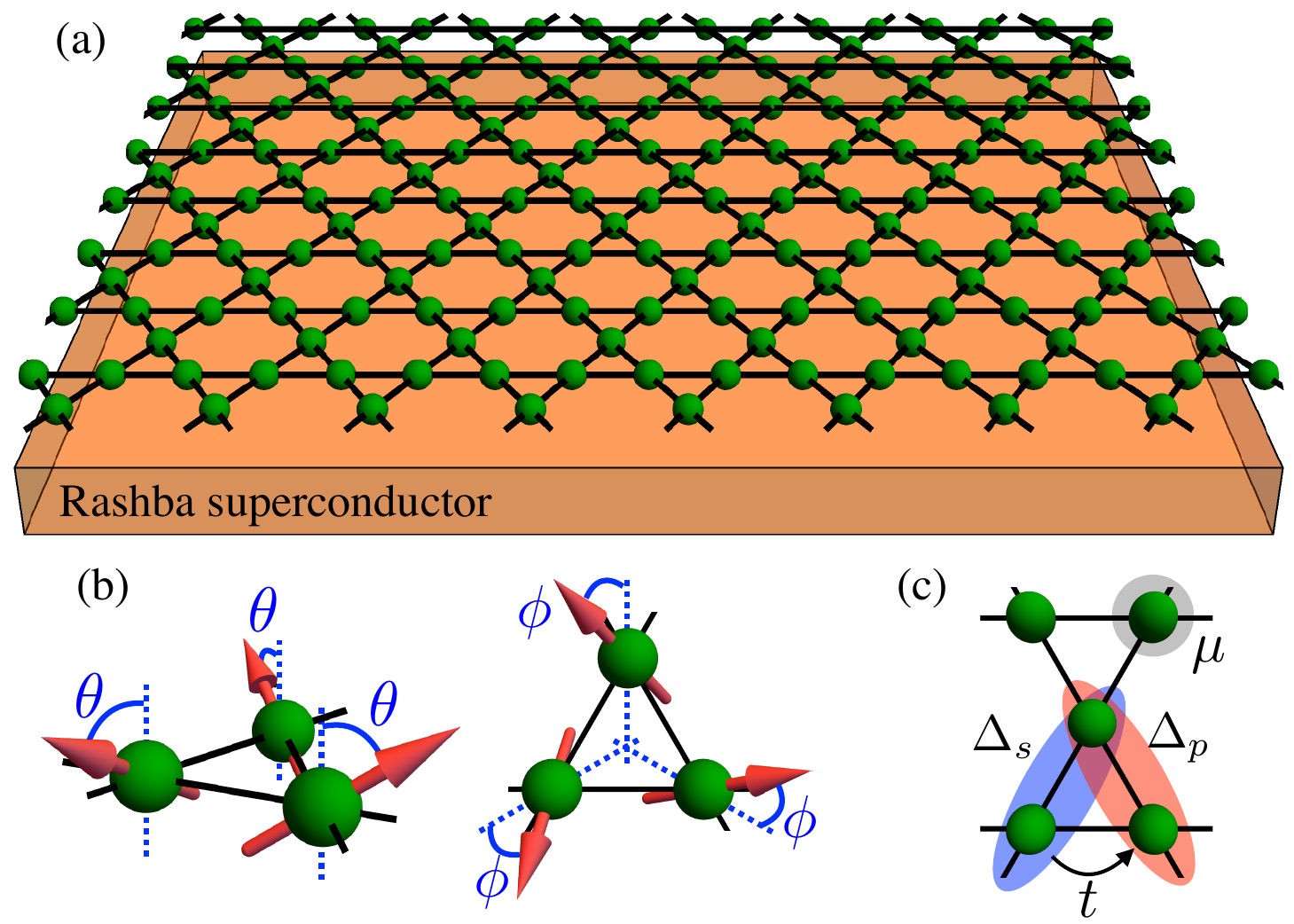}
 \caption{
 (a) Schematic illustration of the hybrid system considered here. The quantum
 anomalous Hall system on a kagome lattice (top layer) is proximity-coupled to
 the Rashba $(s+p)$-wave superconductor (bottom layer).
 (b) Localized spins (red arrows) on a kagome lattice. Their orientations are
 specified by $\theta$ and $\phi$.
 (c) Model parameters appearing in $H$.
 }
 \label{fig:hybrid}
\end{figure}

In this work, we propose coupling to a Rashba superconductor as an alternative 
route to realize TSC in a kagome magnet [Fig.~\ref{fig:hybrid}(a)]. 
A Rashba superconductor lacks 
inversion symmetry, allowing mixing of $s$-wave and $p$-wave 
pairings~\cite{Gorkov01,Zhang08,Sato10,Alicea10,Yu16,Ptok18},
and 
thus induces proximity effects on the kagome side.
We demonstrate that this heterostructure exhibits topological
superconducting phases with odd Bogoliubov-de Gennes (BdG) Chern 
numbers. As an independent characterization of the topological nature, we also 
compute the chiral central charge using the modular-commutator-based 
formula~\cite{Kim22}, which we implement in BdG framework by expressing the 
modular commutator in terms of the Nambu-space correlation matrix. We further 
find that the superconducting proximity effect influences the energetics of the
magnetic ordering, indicating that the preferred magnetic ordering can depend 
on the presence of superconducting pairing.

\section{Model}
In this section, we set up the model. We first introduce the spinful 
Hamiltonian in Sec.~\ref{sec:spinful}, and then derive an effective model in 
the strong-exchange limit in Sec.~\ref{sec:eff}.

\subsection{Spinful Hamiltonian}
\label{sec:spinful}
We consider a system of spinful electrons on a kagome magnet
proximity-coupled to a superconductor. The total Hamiltonian reads
\begin{align}
 H_\tx{hop}+H_\tx{exc}+H_\Delta,
 \label{eq:spinfulHam}
\end{align}
where each term represents the electron hopping, the exchange coupling, and the
proximity-induced pairing, respectively:
\begin{equation}
 \begin{split}
  &H_\tx{hop}
  =-t\sum_{\langle ij\rangle}\sum_{\sigma=\uparrow\downarrow}
  \tilde{c}_{i\sigma}^\dagger\tilde{c}_{j\sigma}
  -\mu\sum_i\sum_{\sigma}
  \tilde{c}_{i\sigma}^\dagger\tilde{c}_{i\sigma},\\
  &H_\tx{exc}
  =-J_HS\sum_{i}\sum_{\sigma\sigma'}
  \tilde{c}_{i\sigma}^\dagger
  \left(\bm{\sigma}\cdot\bm{n}_{i}\right)_{\sigma\sigma'}
  \tilde{c}_{i\sigma'},\\
  &H_\Delta
  =\frac{1}{2}\sum_{ij}
  \left(\tilde{c}_{i\uparrow}^\dagger,\tilde{c}_{i\downarrow}^\dagger\right)
  \left(d_0^{ij}+\bm{d}^{ij}\cdot\bm{\sigma}\right)i\sigma_y
  \left(
  \begin{array}{c}
   \tilde{c}_{j\uparrow}^\dagger \\
   \tilde{c}_{j\downarrow}^\dagger
  \end{array}
  \right)+\tx{h.c.}
 \end{split}
 \label{eq:H-hop-exc-delta}
\end{equation}
Here, $\tilde{c}_{i\sigma}^\dagger$ creates an electron 
with spin $\sigma$ on site $i$, $t$ is the nearest-neighbor (NN) hopping
amplitude, $\mu$ is the chemical potential,
$J_H$ is the strength of the exchange coupling between electrons 
and a localized classical spin with amplitude $S$ and direction $\bm{n}_i$, and
$d_0^{ij}$ and $\bm{d}^{ij}\equiv(d_x^{ij},d_y^{ij},d_z^{ij})$ represent
the singlet and triplet components of proximity-induced pairing between sites
$i$ and $j$. 
We neglect the intrinsic spin-orbit coupling on the kagome side as a
minimal model.

We assume that the localized spins form a three-sublattice magnetic
order on the kagome lattice, sharing a common polar angle $\theta$ while their 
azimuthal angles are related by the $C_3$-rotational symmetry of each upward 
triangle, parameterized by a single azimuthal angle $\phi$, as illustrated in 
Fig.~\ref{fig:hybrid}(b).

Before specifying the form of the pairing, we clarify 
the physical setting considered in this work. We focus on the strong-exchange 
limit $J_HS/t\rightarrow\infty$, where the electron spins are fully
polarized by the local magnetic moments. 
In this strong coupling limit, inversion symmetry forces electrons at 
momenta $\bm{k}$ 
and $-\bm{k}$ in the same band to have parallel spins, and thus a conventional 
$s$-wave superconductor does not induce proximity pairing in the kagome side.

To overcome this limitation, we consider proximity coupling to a 
noncentrosymmetric superconductor, where spin-singlet and spin-triplet pairing
components are mixed. As a representative example, we choose a Rashba 
superconductor;
the singlet component is $s$-wave pairing while the $\bm{d}$-vector for
the triplet component is assumed to take 
$\bm{d}(\bm{k})\propto \bm{k}\times \bm{e}_z$~\cite{Edelshtein89,Frigeri04,Yanase07,Fujimoto07,Bauer12} in the momentum space, where $\bm{e}_z$ is the
unit vector normal to the kagome plane. 
(This $\bm{d}$-vector can be found in a simple Rashba system, see 
Appendix \ref{sec:Rashba}.)
As a minimal setting for such a pairing, we take the following representation 
in the real space:
\begin{align}
 d_0^{ij}
 &=\left\{
 \begin{array}{ll}
  \tilde{\Delta}_s & \tx{for $i=j$}\\
  \Delta_s & \tx{for NN pairs}\\
  0 & \tx{otherwise}\\
 \end{array}
 \right.
 \label{eq:d0}\\
 \bm{d}^{ij}
 &=\left\{
 \begin{array}{ll}
  -i\Delta_p\,\bm{e}_{ij}\times\bm{e}_z & \tx{for NN pairs}\\
  0 & \tx{otherwise},\\
 \end{array}
 \right.
 \label{eq:dvector}
\end{align}
where $\tilde{\Delta}_s$, $\Delta_s$, and $\Delta_p$ 
are model parameters that represent the proximity-induced pairing
amplitudes,
and $\bm{e}_{ij}$ is the unit vector along the bond from site $j$ to $i$. 

Although the Rashba superconductor preserves time-reversal symmetry, it lacks
inversion symmetry and thus enables proximity pairing on the kagome side.
On the other hand, the kagome side breaks time-reversal symmetry due to 
magnetic ordering. Thus, the hybrid system combines proximity-induced
pairing with broken time-reversal symmetry, allowing for TSC.

\subsection{Strong-exchange limit}
\label{sec:eff}
Here we derive an effective model in the limit $J_HS/t\rightarrow\infty$, 
where,
as mentioned above, electrons completely align with the direction $\bm{n}_i$ of
the localized spin at each site. Hereafter, all results are obtained in this 
limit.
The creation 
operator of such an electron is
\begin{align}
 c_i^\dagger\equiv
 (\tilde{c}^\dagger_{i\uparrow},\tilde{c}^\dagger_{i\downarrow})\bm{\chi}_i.
\end{align}
Here, $\bm{\chi}_i$ is an eigenvector of $\bm{\sigma}\cdot\bm{n}_i$ with an 
eigenvalue $+1$:
\begin{align}
 \bm{\chi}_{i}
 =
 \left(
 \begin{array}{c}
  e^{-i\phi_i/2}\cos\left(\theta_i/2\right)\\
  e^{i\phi_i/2}\sin\left(\theta_i/2\right)
 \end{array}
 \right),
 \label{eq:chi}
\end{align}
where $(\theta_i,\phi_i)$ are the polar coordinates of $\bm{n}_i$.
The projection onto the spin-polarized subspace is performed by the
transformation 
$(\tilde{c}^\dagger_{i\uparrow},\tilde{c}^\dagger_{i\downarrow})\rightarrow
c_i^\dagger\bm{\chi}_i^\dagger$ in Eq.~\eqref{eq:H-hop-exc-delta},
yielding a single (spinless) fermion model in which the itinerant electron spin
is polarized along the local-moment direction $\bm{n}_i$.
We thereby obtain an effective Hamiltonian as 
\begin{align}
 &H
 =-t\sum_{\langle ij\rangle}
 c_i^\dagger\bm{\chi}_i^\dagger
 \bm{\chi}_j c_j
 -(\mu+J_HS)\sum_{i}c_i^\dagger c_i\non
 &
 +\frac{1}{2}\sum_{i}\left(
 \tilde{\Delta}_s
 c_i^\dagger\bm{\chi}_i^\dagger
 i\sigma_y\bm{\chi}_i^*c_i^\dagger
 +\tx{h.c.}
 \right)
 \non
 &+\frac{1}{2}\sum_{\langle ij\rangle}
 \left(
 c_i^\dagger\bm{\chi}_i^\dagger
 \left(\Delta_s
 -i\Delta_p\left(\bm{e}_{ij}\times\bm{e}_z\right)\cdot\bm{\sigma}\right)
 i\sigma_y\bm{\chi}_j^*c_j^\dagger+\tx{h.c.}
 \right).
 \label{eq:effHam}
\end{align}
Here, the third term describing the on-site pairing vanishes due to the 
Pauli principle, $c_i^\dagger c_i^\dagger=0$.
Note that the Hamiltonian $H$ still depends on the directions of the 
localized spins through $\bm{\chi}_i$ [Eq.~\eqref{eq:chi}], appearing in the
hopping and pairing terms.
Hereafter, we redefine the chemical potential $\mu$ to include the constant 
shift $J_HS$
(i.e., $J_H$ no longer appears below). 
In the BdG representation in the momentum space, the Hamiltonian 
becomes
\begin{equation}
 \begin{split}
  H
  &=\frac12\sum_{\bm{k}}\left(
  \bm{c}^\dagger(\bm{k})
  h(\bm{k})
  \bm{c}(\bm{k})+\tx{Tr\,}h_\tx{hop}(\bm{k})\right),\\
  h(\bm{k})
  &=\left(
  \begin{array}{cc}
   h_\tx{hop}(\bm{k}) & \Delta(\bm{k})\\
   \Delta^\dagger(\bm{k}) & -h_\tx{hop}^*(-\bm{k})
  \end{array}
  \right),
  \label{eq:Heff}
 \end{split}
\end{equation}
where $h_\tx{hop}(\bm{k})$ and 
$\Delta(\bm{k})$ are $3\times3$ matrices, and $\bm{c}^\dagger(\bm{k})
=\left(c^\dagger_1(\bm{k}),c^\dagger_2(\bm{k}),c^\dagger_3(\bm{k}),
c_1(-\bm{k}),c_2(-\bm{k}),c_3(-\bm{k})\right)$ are the creation operators and 
the subscripts indicate sublattice. The explicit forms of $h_\tx{hop}(\bm{k})$ 
and $\Delta(\bm{k})$ are given in Appendix \ref{sec:BdGelements}.

As summarized in Figs.~\ref{fig:hybrid}(b) and (c), the model parameters at 
this stage are
\begin{align*}
 t, \mu, \Delta_s, \Delta_p, \theta, \phi.
\end{align*}
Without loss of generality, we set $\Delta_s$ and $\Delta_p$ to be positive
real numbers, under an assumption that the pairing term preserves time-reversal
symmetry~\cite{Deltasp}. We also restrict the parameter space to 
$0\leq\theta\leq\pi/2$ and $0\leq\phi\leq\pi$, since the other regions are
related to this one by unitary transformations~\cite{thetaphi}. 

\section{Topological characterization}
Our system belongs to symmetry class D~\cite{Altland97,Ryu10}, 
since time-reversal symmetry is
broken by the magnetic ordering while particle-hole symmetry is preserved.
To topologically characterize emergent phases, we calculate the BdG Chern
number $\mathcal{N}$ and the chiral central charge $c_-$. Although the relation
$\mathcal{N}/2=c_-$ holds in our case, the two quantities arise from
different formulations, and calculating both provides a useful consistency
check. Below, we describe each topological invariant.

\subsection{BdG Chern number $\mathcal{N}$}
The BdG Chern number $\mathcal{N}$~\cite{Qi10} is the superconducting analogue 
of the ordinary Chern number for Chern 
insulators~\cite{Thouless82,Kohmoto85,Kudo19}.
The value of $|\mathcal{N}|$ corresponds to the number of chiral 
Majorana edge modes, and the sign of $\mathcal{N}$ specifies their propagation 
direction.
For the BdG Hamiltonian matrix $h(\bm{k})$ in Eq.~\eqref{eq:Heff}, let us first
define the non-Abelian Berry connection one-form as
\begin{align}
 A
 =\Psi^\dagger d\Psi
\end{align}
where $\Psi=\left(\bm{\psi}_1, \bm{\psi}_2,\bm{\psi}_3\right)$ and
$\bm{\psi}_i$ is the lowest $i$th eigenvector of $h(\bm{k})$. Note that
$A$ is a $3\times3$ matrix.
The field strength two-form and the BdG Chern number are
\begin{equation}
 \begin{split}
  F&=dA+A\wedge A,\\
  \mathcal{N}&=\frac{1}{2\pi i}\int_{BZ}\tx{Tr\,}F,
 \end{split}
\end{equation}
where the integral is taken over the Brillouin zone. For the numerical 
calculations, we adopt the discretized Brillouin-zone method~\cite{Fukui05}.

\begin{figure}[t]
\includegraphics[width=\columnwidth]{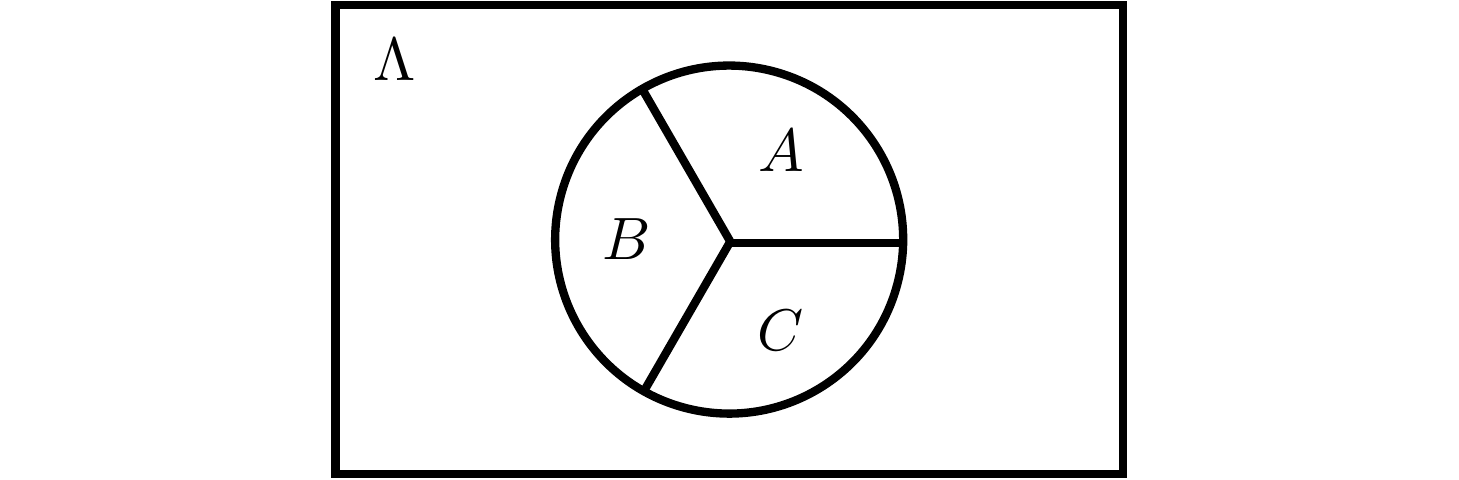}
 \caption{
 Partition of subsystems $A$, $B$, and $C$ in the bulk $\Lambda$.
 }
 \label{fig:modular}
\end{figure}
\subsection{Chiral central charge $c_-$}
The chiral central charge $c_-$ is a topological invariant, which
appear as the quantized thermal Hall conductance at low 
temperature~\cite{Kane97,Kitaev06,Read00,Nomura12}. Recently, it has
been shown that $c_-$ can be extracted directly from a single many-body ground 
state as~\cite{Kim22,Kim22b,Fan22,Fan23,Fan23b}
\begin{align}
 J(A,B,C)=\frac{\pi}{3}c_-.
\end{align}
Here $J(A,B,C)$ is the modular commutator:
\begin{align}
 J(A,B,C)=i\tx{Tr\,}\left(\rho^{(ABC)}\left[K^{(AB)},K^{(BC)}\right]\right),
 \label{eq:modJ}
\end{align}
where $A$, $B$, and $C$ represent subsystems as shown in 
Fig.~\ref{fig:modular}, $AB$ denotes the union $A\cup B$,
$\rho^{(X)}=\tx{Tr}_{\bar{X}}\,\rho$ is
the reduced density matrix for a subsystem $X$, and 
$K^{(X)}=-\tx{log\,}\rho^{(X)}$ is the modular 
(entanglement) Hamiltonian. Here, $\rho=\ket{G}\bra{G}$ is the ground 
state density matrix and $\tx{Tr}_{\bar{X}}$ denotes the trace over the 
complement of a subsystem $X$.
In Eq.~\eqref{eq:modJ}, the tensor products with the identity matrix are 
suppressed for simplicity; e.g., $K^{(AB)}=K^{(AB)}\otimes1^{(C)}$.

In BdG systems, the expression of $J(A,B,C)$ can be simplified using Peschel's
formula~\cite{Peschel03} with the Majorana representation~\cite{Zou22}.
Instead, we here develop an simple expression obtained from the Nambu-space 
correlation matrix.

The modular Hamiltonian $K^{(X)}$ for our BdG Hamiltonian $H$
takes the form of fermion bilinear~\cite{Peschel03}:
\begin{align}
 K^{(X)}
 =\frac12\sum_{ij}\left(c_i^\dagger,c_i\right)k^{(X)}_{ij}
 \left(
 \begin{array}{c}
  c_j \\
  c_j^\dagger
 \end{array}
 \right),
 \label{eq:modH}
\end{align}
where $k^{(X)}_{ij}$ is a $2\times2$ matrix, and we denote the full block 
matrix by $k^{(X)}$.
Substituting Eq.~\eqref{eq:modH} into Eq.~\eqref{eq:modJ}, the modular 
commutator becomes
\begin{align}
 J(A,B,C)=-\frac{i}{2}\tx{Tr}\left(
 \left[k^{(AB)},k^{(BC)}\right]\mathcal{C}^{(ABC)}\right),
 \label{eq:JhhC}
\end{align}
where $\mathcal{C}^{(X)}$ is a block matrix, whose $2\times2$ subblocks are 
the correlation matrices between sites in $X$:
\begin{align}
 \mathcal{C}_{ij}=\bra{\Omega}
 \left(
 \begin{array}{c}
  c_i \\
  c_i^\dagger
 \end{array}
 \right)
 \left(c_j^\dagger,c_j\right)
 \ket{\Omega}\ 
 \left(i,j\in X\right).
\end{align}
Here, $\ket{\Omega}$ is the quasiparticle vacuum state of the BdG Hamiltonian 
$H$. We note 
that the block matrix $k^{(X)}$ is also obtained from the correlation matrix
via~\cite{Oliveira14,Kudo24}
\begin{align}
 k^{(X)}=-\ln\left[\frac{1-\mathcal{C}^{(X)}}{\mathcal{C}^{(X)}}\right].
 \label{eq:modular-k}
\end{align}
This implies that the modular commutator $J(A,B,C)$ can be evaluated 
solely from the correlation matrix $\mathcal{C}^{(X)}$. 
In our numerical calculations of $c_-$, 
we use Eq.~\eqref{eq:JhhC} with Eq.~\eqref{eq:modular-k}.

\begin{figure}[t]
\includegraphics[width=\columnwidth]{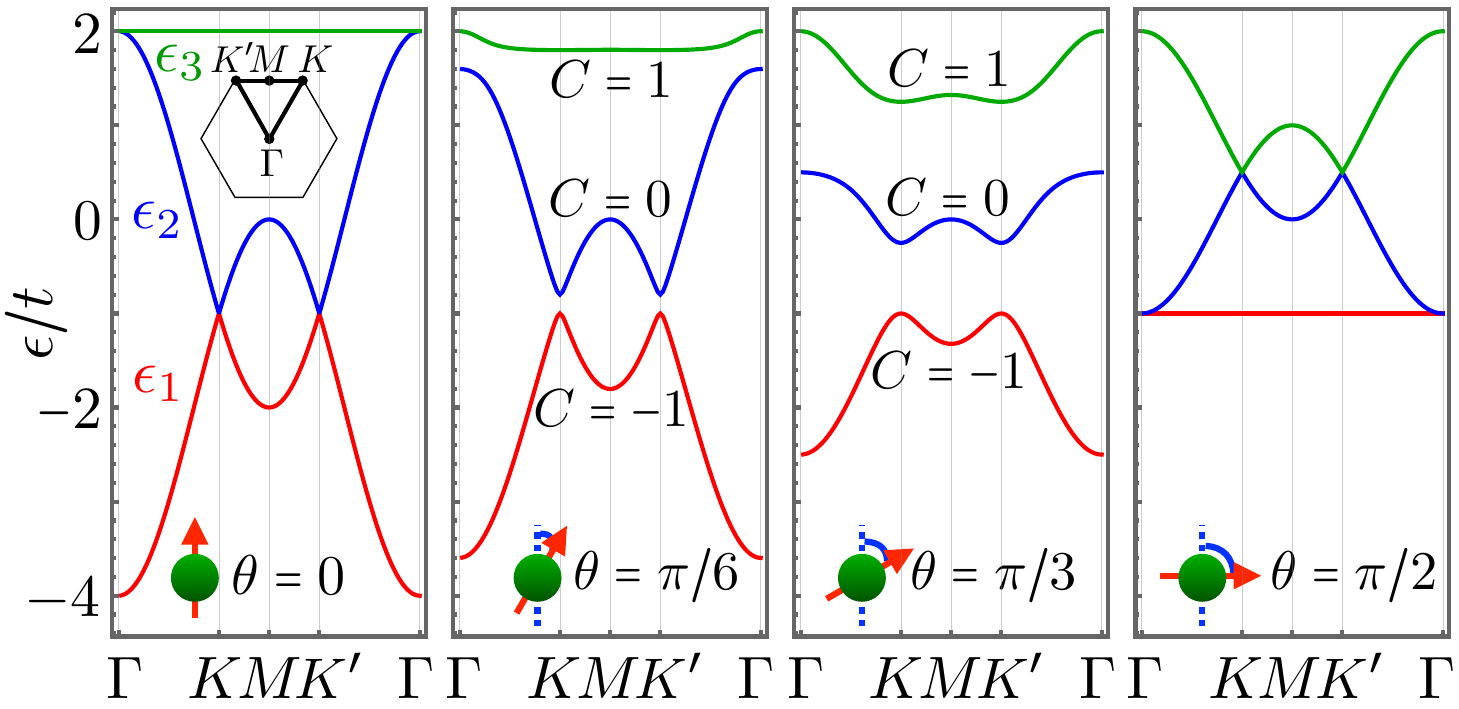}
 \caption{
 Energy bands of $h_\tx{hop}$ for the several polar angles $\theta$. Each band 
 is denoted by $\epsilon_{n}$ with $n=1,2,3$.
 All three bands at $\theta\neq0,\pi/2$ are isolated, and their band Chern
 numbers $C$ are $-1$, 0, 1 from the bottom, respectively.
 }
 \label{fig:band}
\end{figure}
\begin{figure}[t]
\includegraphics[width=\columnwidth]{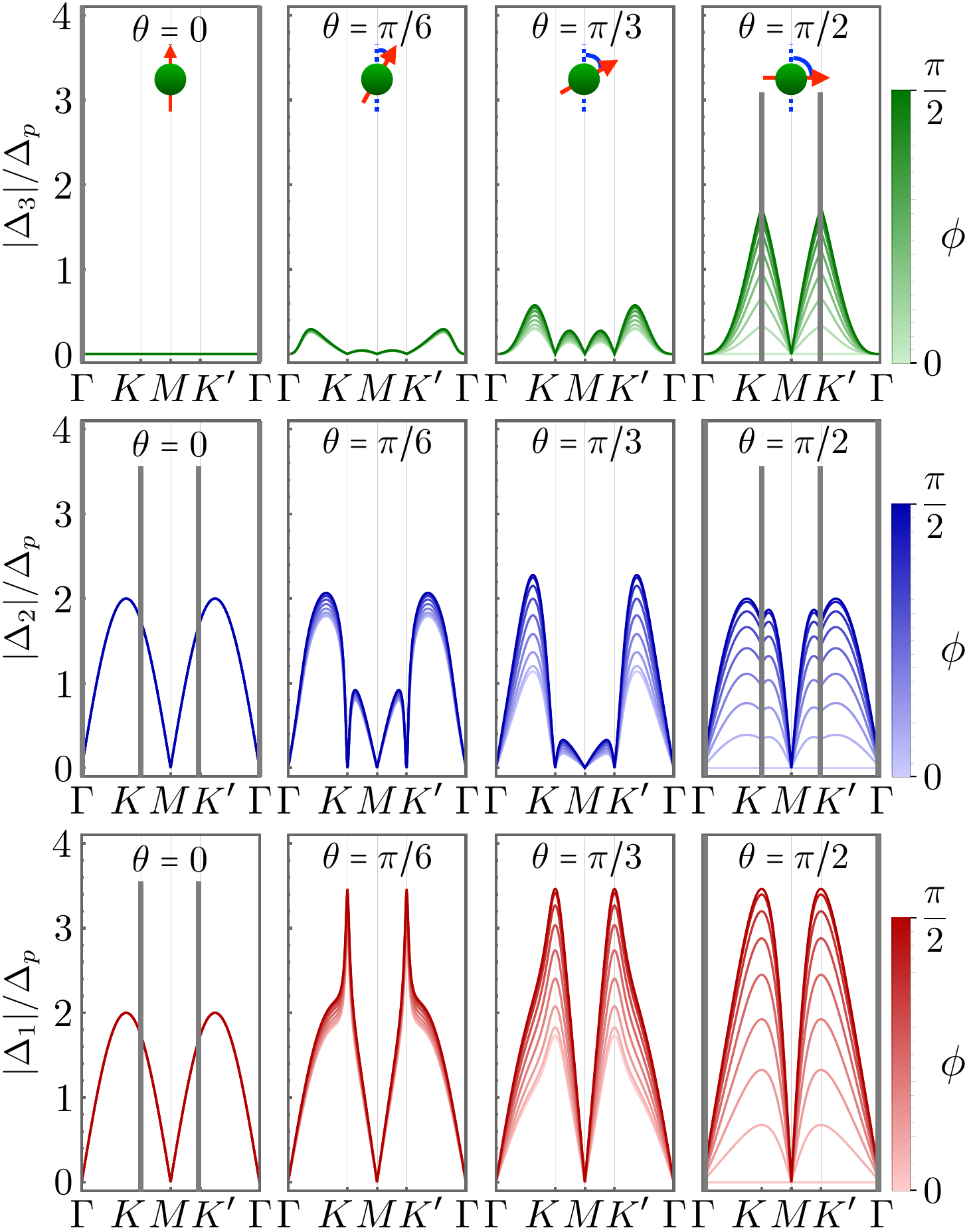}
 \caption{
 Pair potential projected onto each band. Each row
 represents a different band ($\Delta_n$ corresponds to $\epsilon_n$ in 
 Fig.~\ref{fig:band}). From left to right, the polar angle $\theta$ is varied 
 as indicated.
 The color intensity encodes $\phi$.
 We scale $\Delta_n(\bm{k})$ by $\Delta_p$ since the effect of the $s$-wave 
 pairing is not included. The gray vertical bars indicate the momenta at which
 band touchings occur, where $\Delta_n$ is ill-defined.
 We focus on $0\leq\phi\leq\pi/2$ since we have numerically verified that 
 $|\Delta_n(\bm{k})|$ with $\phi$ and $\pi-\phi$ are identical.
 }
 \label{fig:pair-potential}
\end{figure}
\section{Numerical results}
Now we numerically demonstrate the emergence of topological superconducting
phases in our hybrid system composed of a kagome magnet and a Rashba 
superconductor. We evaluate basic properties of the BdG system, whose technical
details are summarized in Appendix \ref{sec:basic}.

\subsection{Quantum anomalous Hall system}
Before discussing the BdG problem, let us briefly review the quantum 
anomalous Hall effect realized in kagome magnets~\cite{Ohgushi00}.
Figure~\ref{fig:band} shows the energy spectra of $h_\tx{hop}$
for the several polar angles $\theta$ (the spectra are independent of the
azimuthal angle $\phi$).
For $\theta\neq0,\pi/2$, all three bands are isolated, and their band Chern
numbers $C$ are $-1$, 0, 1 from the bottom, respectively. The $\theta=0$ and 
$\pi/2$ spectra become identical after flipping the energy axis 
and applying a suitable energy rescaling. 

\subsection{Projected pair potential}
Let us now move onto the BdG problem.
To quantify the proximity effect, we compute the pair potentials 
projected onto each band:
\begin{align}
 \Delta_n(\bm{k})\equiv\bm{u}_n^\dagger(\bm{k})\Delta(\bm{k})
 \bm{u}_n^*(-\bm{k}),
\end{align}
where $\Delta(\bm{k})$ is the off-diagonal part of the BdG
Hamiltonian [see Eq.~\eqref{eq:Heff}], and $\bm{u}_n$ is the eigenvector of
the energy band $\epsilon_n$ (Fig.~\ref{fig:band}).
Figure~\ref{fig:pair-potential} shows $|\Delta_n(\bm{k})|$ for several 
$\theta$ and $\phi$. As mentioned in the introduction, 
because $\bm{u}_n(\bm{k})$ and $\bm{u}_n(-\bm{k})$ have parallel spin
directions due to the inversion symmetry of $h_\tx{hop}$, the $s$-wave 
component does not contribute to $\Delta_n(\bm{k})$; hence $\Delta_n(\bm{k})$ 
is scaled only by $\Delta_p$ in the figure.
(The intraband $s$-wave pairing can appear if the $s$-wave pair potential is 
modulated to break inversion symmetry~\cite{Chaou25}, as discussed in 
Appendix \ref{sec:Chaou-potential}.)
\begin{figure}[t]
\includegraphics[width=\columnwidth]{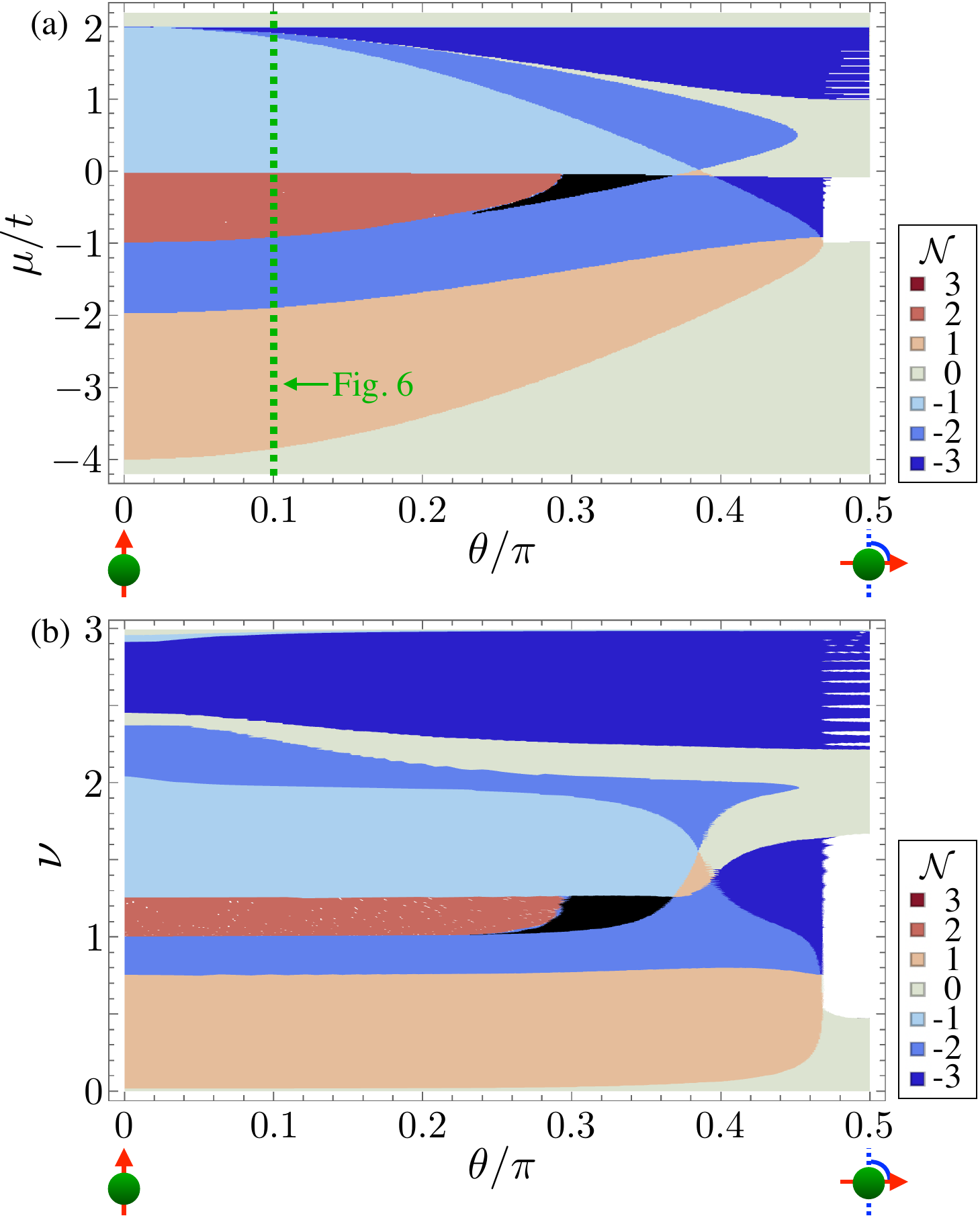}
 \caption{
 BdG Chern number $\mathcal{N}$ as a function of the polar angle $\theta$ and
 (a) the chemical potential $\mu$ or (b) the band filling $\nu$. Black 
 regions represent $|\mathcal{N}|\geq4$. 
 White regions correspond to parameters
 where quasiparticle energy becomes negative (see Appendix~\ref{sec:basic}
 for more details).
 This occurs because inversion symmetry is absent in our system.
 Other parameters are set to be $\Delta_s/t=\Delta_p/t=0.1$ and 
 $\phi=\pi/2$.
 }
 \label{fig:PD}
\end{figure}

The $\phi$-dependence of $\Delta_n$ results from anisotropy of the $p$-wave 
pairing combined with the kagome lattice structure. The results indicate that
$|\Delta_n|$ increases as $\phi\to\pi/2$ as a general feature. At $\phi=\pi/2$,
$\Delta_1$ is enhanced for large $\theta$, which will be exploited below to 
control the magnetic ordering of the system.

Because of $\Delta_n(\bm{k})=-\Delta_n(-\bm{k})$, nodes necessarily appear at 
the $\Gamma$ and $M$ points. Additional nodes occur at $K$ and $K'$ for
$\Delta_{2}$ and $\Delta_{3}$. These nodal structures naturally allow for 
TSC with odd BdG Chern numbers $\mathcal{N}$, as demonstrated below.
Most nodes exhibit 
linear dispersion, while $\Delta_3(\bm{k})$ vanishes as $k^3$ at $\Gamma$, 
indicating an $f$-wave pairing structure associated with $|\mathcal{N}|=3$.

\begin{figure}[t]
\includegraphics[width=\columnwidth]{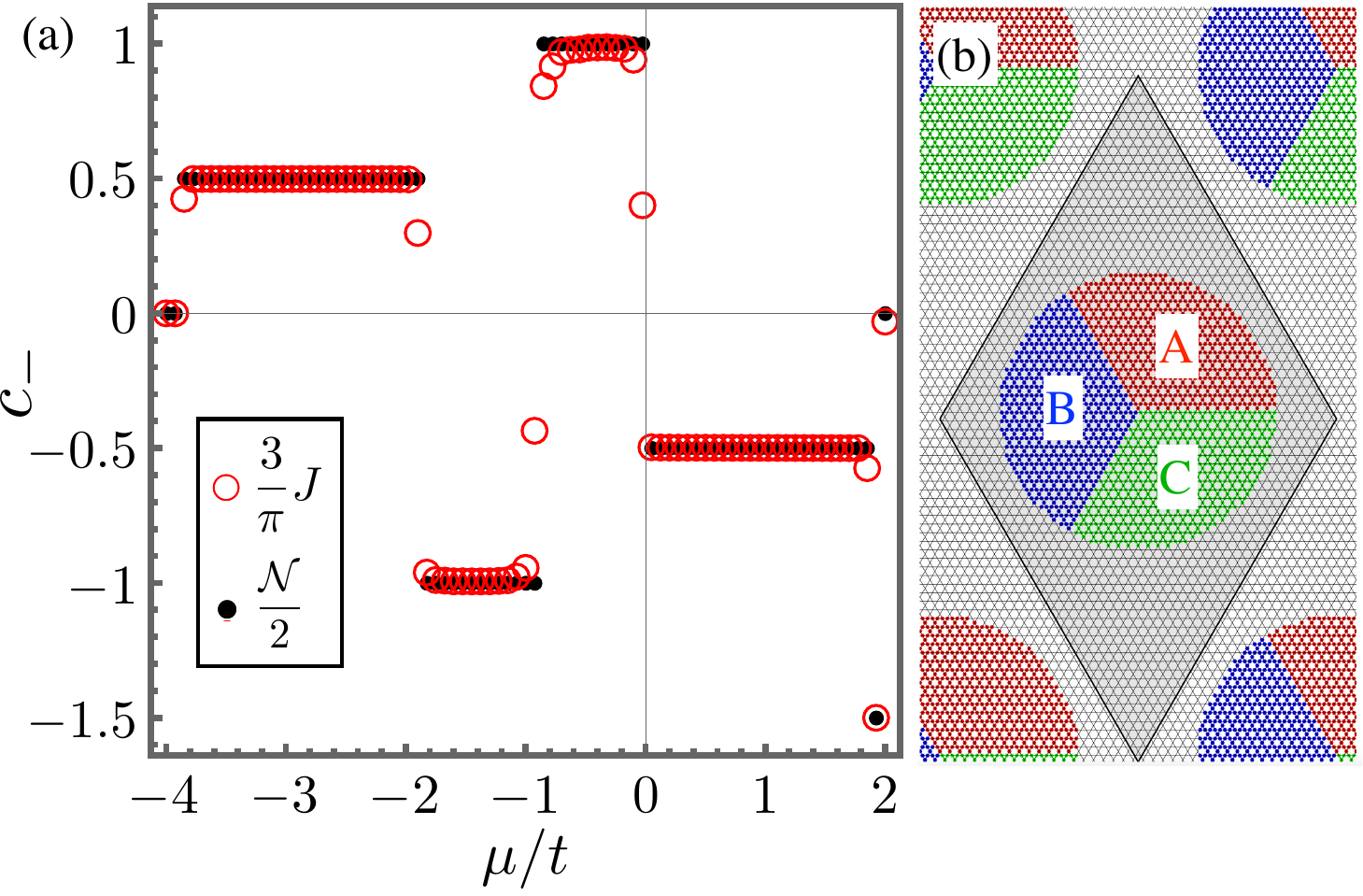}
 \caption{
 (a) Chiral central charge $c_-$ as a function of the chemical potential $\mu$,
 evaluated along the green line in Fig.~\ref{fig:PD}(a). Red circles 
 represent the values obtained from the modular commutator $J(A,B,C)$ via
 Eq.~\eqref{eq:JhhC}, while black dots show one-half of the BdG Chern number 
 $\mathcal{N}$.
 (b) Subsystems $A$, $B$, and $C$ used for calculating $J(A,B,C)$. The system 
 size is $40\times40$ unit cells (gray rhombus), and the radius of the 
 subsystem $(A,B,C)$ is $\sim14$ in units of the lattice constant.
 }
 \label{fig:ccc}
\end{figure}
\subsection{Phase diagram}
To investigate TSC, we plot in 
Fig.~\ref{fig:PD}(a) the BdG Chern number $\mathcal{N}$ as a function
of $\theta$ and the chemical potential $\mu$. Here, we set
$\Delta_s/t=\Delta_p/t=0.1$ and $\phi=\pi/2$. Figure~\ref{fig:PD}(b) 
presents the same phase diagram, but with the vertical axis 
rearranged with respect to the band filling $\nu$, obtained by calculating the 
expectation value of the particle number. In particular, 
topological superconducting phases with odd $\mathcal{N}$ spread broadly over 
the regions of fractional filling. 
This reflects the fact that the nodal structure of the pair potential
shown in Fig.~\ref{fig:pair-potential} affects the BdG band topology only at 
fractional filling.
As mentioned in the previous section,
a phase with $\mathcal{N}=-3$ emerges around $\nu\lesssim3$. 
The topological superconducting phases identified here are expected to
be robust against perturbations such as disorder or intrinsic spin-orbit
coupling on the kagome side, as long as the bulk gap remains open.

We here discuss the robustness of the phase diagram against deviations 
of $\bm{d}(\bm{k})$ from $\bm{k}\times\bm{e}_z$. If the Rashba-type 
superconducting pairing originates from an interface that breaks reflection
about the plane, possible deviations are 
longer-range contributions such as next-nearest-neighbor pairings, which are
expected to be strongly suppressed due to the exponential decay with respect
to the distance. If Rashba-type 
superconducting pairing originates from an intrinsically 
noncentrosymmetric superconductor, the deviations may differ depending on the
material.
In general, the emergent topological superconducting phases are
determined by the location and the topological structure of the nodes in the
projected pair potential (Fig.~\ref{fig:pair-potential}). These nodes are 
pinned to high-symmetry points of the Brillouin zone, namely, $\Gamma$, $M$, 
$K$, and $K'$: the nodes at $\Gamma$ and $M$ come from the symmetry relation
$\Delta_n(\bm{k})=-\Delta_n(-\bm{k})$, which reflects the odd-parity nature
of spin-triplet pairing, and therefore persist regardless of the detailed form
of deviations. (In our case, deviations only with spin-triplet pairing are 
allowed due to the strong-exchange limit.) The nodes at $K$ and $K'$, on the
other hand, are expected to originate from symmetric or topological properties
of the band wavefunctions and the present form of $\bm{d}(\bm{k})$. 
Consequently, the emergent phases  in Fig.~\ref{fig:PD} may change depending on
whether the nodes at $K$ and $K'$ survive such deviations.

In Fig.~\ref{fig:ccc}(a), we also evaluate the chiral central charge $c_-$ 
from the modular commutator $J(A,B,C)$ along the green line in 
Fig.~\ref{fig:PD}(a). 
Here, the system size is $40\times40$ unit cells. The dependence
on the system size is discussed in Appendix~\ref{appx:finite-size}.
The resulting values show perfect agreement with one-half
of the BdG Chern number $\mathcal{N}$ and clearly capture the phase 
transitions. This demonstrates that the modular-commutator-based evaluation of 
$c_-$ provides a practical and reliable characterization of topological 
superconducting phase.
Figure~\ref{fig:ccc}(b) shows the configuration of subsystems
$A$, $B$, and $C$ used for calculating $J(A,B,C)$.

\subsection{Magnetic ordering minimizing the quasiparticle vacuum energy}
So far, $\theta$ and $\phi$ (directions of localized spins) have been treated 
as given parameters. We now regard them as variational degrees of freedom and 
determine the energy-minimizing configuration.

Since $\phi=\pi/2$ yields the most enhanced projected pair potentials 
[Fig.~\ref{fig:pair-potential}], which implies lower ground state energy, 
we fix $\phi$ at this value and seek the 
energy-minimizing $\theta$, denoted by $\theta_\tx{min}$, for a given $\nu$. 
Figure~\ref{fig:Eg}(a) shows
the quasiparticle vacuum energy $E_g$ (defined in Appendix~\ref{sec:basic}) at 
$\nu=0.6$ as a function 
of $\theta$ for several values of the pairing amplitude. Here we set
$\Delta_s=\Delta_p\equiv\Delta_{s,p}$ for simplicity.
For $\Delta_{s,p}=0$, $\theta_\tx{min}$ is determined solely by the kinetic 
energy and is found to be zero (the same behavior is observed for
all fillings $0<\nu<3$). As $\Delta_{s,p}$ increases, $\theta_\tx{min}$ 
switches sharply from 0 to $\pi/2$ as seen in Fig.~\ref{fig:Eg}(b). 
This transition is necessarily discontinuous: as shown in 
Fig.~\ref{fig:Eg}(a), $E_g$ as a function of $\theta$ has local minima only at
$\theta=0$ and $\theta=\pi/2$ and thus $\theta_\tx{min}$ can take only these
two values. This 
transition reflects the competition between kinetic energy (favoring 
$\theta=0$) and pairing (favoring $\theta=\pi/2$ as confirmed in
Fig.~\ref{fig:pair-potential}), demonstrating that the magnetic ordering can be
controlled by the superconducting proximity effect.

\begin{figure}[t]
\includegraphics[width=\columnwidth]{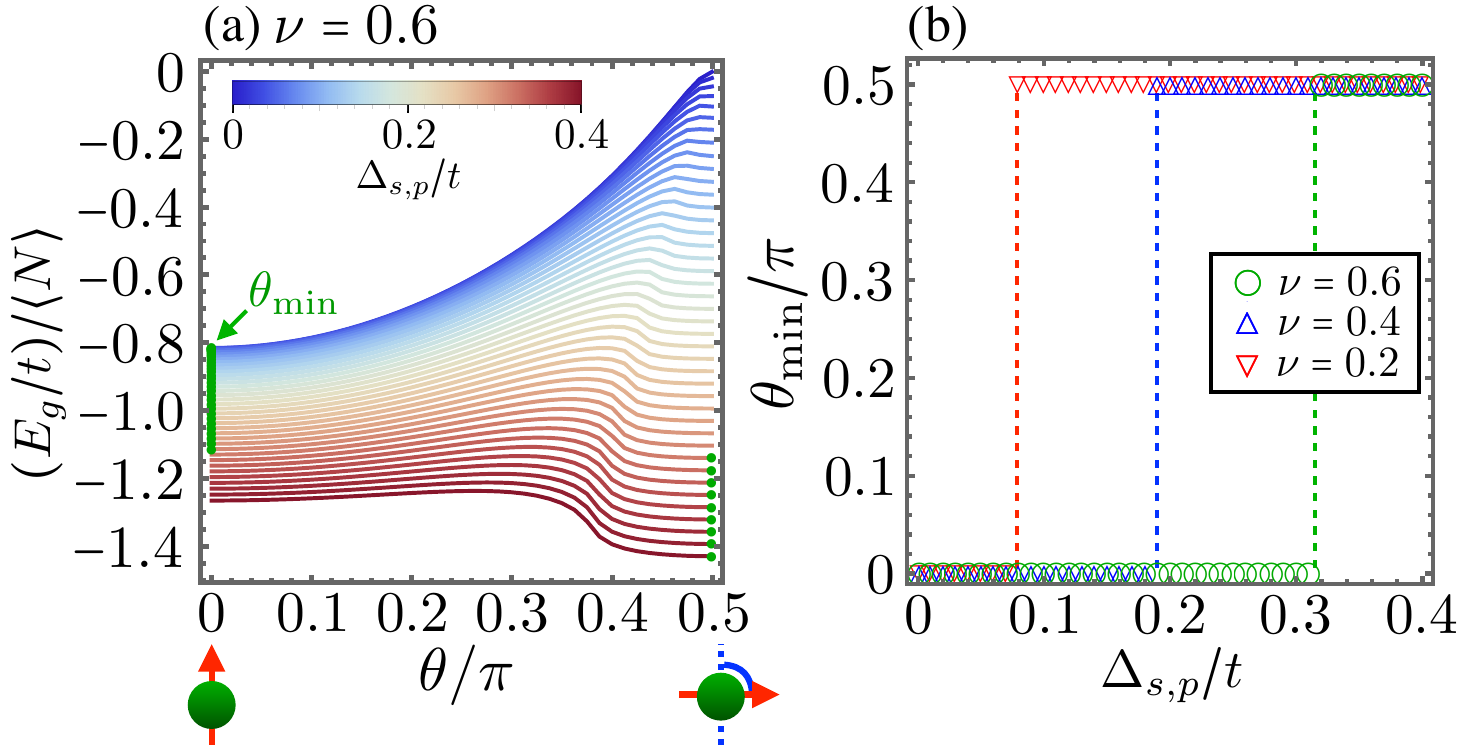}
 \caption{
 (a) Quasiparticle vacuum energy $E_g$ divided by the expectation value of the 
 particle
 number $\langle N\rangle$ for $\nu=0.6$. The color indicates the pairing 
 amplitude $\Delta_s=\Delta_p\equiv\Delta_{s,p}$. The green dots represent the 
 polar angle $\theta_\tx{min}$ that minimizes $E_g$ at each $\Delta_{s,p}$.
 (b) $\theta_\tx{min}$ as a function of $\Delta_{s,p}$ for several filling
 factors. The system size is $30\times30$ unit cells in both panels.
 }
 \label{fig:Eg}
\end{figure}

This behavior persists for other fillings, as shown in Fig.~\ref{fig:Eg}(b). 
The critical value of $\Delta_{s,p}$ increases with $\nu$, because 
pairing occurs only near the Fermi surface whereas kinetic energy involves all 
electrons, reducing the relative impact of pairing at large $\nu$.
We confirm that $\theta_\tx{min}$ remains zero at $\nu=1.6$ and 2.6
within $0\leq\Delta_{s,p}/t\leq0.4$, although the data are not shown in the 
figure to avoid overcrowding.

\section{Summary}
We have demonstrated that topological superconductivity emerges on kagome
magnets coupled to Rashba superconductors. The resulting
phases are characterized by the projected pair potentials, the BdG Chern 
number, and the chiral central charge. 
By treating the spin orientations as
a variational degree of freedom, we have shown that the superconducting
proximity effect can select and thereby control the underlying magnetic 
ordering.

From an experimental perspective, superconductors with antisymmetric
spin-orbit coupling can be realized
in noncentrosymmetric superconductors, such as CePt$_3$Si~\cite{Bauer04}, 
CeRhSi$_3$~\cite{Kimura05}, CeIrSi$_3$~\cite{Sugitani06}, and 
UIr~\cite{Akazawa04}. Even in centrosymmetric materials, Rashba pairing can be 
induced in superconducting heterostructures with interfacial inversion-symmetry
breaking. Together with kagome materials, our results may be accessible in
realistic experimental settings.

In our work, electron-electron interactions on the kagome side are
ignored. In Landau-level systems, the interplay between repulsive
interactions and proximity-induced pairing can yield exotic topological
superconducting states~\cite{Kudo25}. It would be interesting to generalize 
our work in this direction.

\begin{acknowledgments}
 K.K. thanks Kohtaro Kato for helpful discussions about the modular commutator.
 The work is supported in part by JSPS KAKENHI Grant nos. 
 JP23K19036, 
 JP24K06926, 
 JP25K17318, 
 JP25H01250, 
 JP25H00613. 
\end{acknowledgments}

\appendix

\section{Cooper pairing in a Rashba system}
\label{sec:Rashba}
Here, we show that Cooper pairing in a simple Rashba system is described by 
$\bm{d}(\bm{k})\propto \bm{k}\times \bm{e}_z$. 
As a minimal model capturing this structure, we consider the Hamiltonian of an 
electron gas with Rashba spin-orbit coupling in two dimensions:
 \begin{align}
  H=\frac{p^2}{2m}-\alpha_R\left(\bm{\sigma}\times\bm{p}\right)_z,
 \end{align}
where $\alpha_R$ is the Rashba coupling strength. The eigenstates are given by
\begin{align}
 H\ket{\bm{k},\pm}
 =\left[\frac{\hbar^2k^2}{2m}
 \pm\alpha_R\hbar k\right]
 \ket{\bm{k},\pm},
\end{align}
where $\bm{k}$ is the wave vector and
$-(\bm{\sigma}\times\hat{\bm{k}})_z\ket{\pm}=\pm\ket{\pm}$ with
$\hat{\bm{k}}=\bm{k}/|\bm{k}|$. Using 
$-(\bm{\sigma}\times\hat{\bm{k}})_z=\hat{k}_x\sigma_y-\hat{k}_y\sigma_x$, 
we 
have
\begin{align}
 \ket{\bm{k},\pm}
 &=\frac{1}{\sqrt{2}}\left(\ket{\bm{k},\uparrow},\ket{\bm{k},\downarrow}\right)
 \left(
 \begin{array}{c}
  e^{-i\theta_{\bm{k}}/2} \\
  \pm e^{i\theta_{\bm{k}}/2}
 \end{array}
 \right)
 \label{eq:pm}\\
 e^{i\theta_{\hat{\bm{k}}}}
 &=-\hat{k}_y+i\hat{k}_x,
 \label{eq:thetak}
\end{align}
where $\ket{\uparrow}$ and $\ket{\downarrow}$ are eigenstates of $\sigma_z$ 
with the eigenvalues $1$ and $-1$, respectively.

Now, we consider pairing between the time-reversal partners
$\ket{\bm{k},+}$ and $\ket{-\bm{k},-}$. The pair creation 
operator is decomposed into spin-singlet and spin-triplet components as
\begin{align}
 c^\dagger_{\bm{k},+}c^\dagger_{-\bm{k},-}
 &=\frac{1}{2}
 \left(
 c^\dagger_{\bm{k},+}c^\dagger_{-\bm{k},-}
 -c^\dagger_{\bm{k},-}c^\dagger_{-\bm{k},+}\right)\non
 &\qquad
 +\frac{1}{2}
 \left(
 c^\dagger_{\bm{k},+}c^\dagger_{-\bm{k},-}
 +c^\dagger_{\bm{k},-}c^\dagger_{-\bm{k},+}\right)\non
 &\equiv
 A_\tx{sin}+A_\tx{tri}.
\end{align}
Using Eq.~\eqref{eq:pm}, each component is expressed as
\begin{align}
 A_\tx{sin/tri}
 &=\frac{1}{2}
 \left(
 c^\dagger_{\bm{k},+}c^\dagger_{-\bm{k},-}
 \mp c^\dagger_{\bm{k},-}c^\dagger_{-\bm{k},+}\right)\non
 &=\frac{1}{2}\times
 \left\{
 \begin{array}{l}
  \left(
   -c^\dagger_{\bm{k}\uparrow}c^\dagger_{-\bm{k}\downarrow}
   +c^\dagger_{\bm{k}\downarrow}c^\dagger_{-\bm{k}\uparrow}
  \right)
  \\
  \left(
   e^{-i\theta_{\hat{\bm{k}}}}
   c^\dagger_{\bm{k}\uparrow}c^\dagger_{-\bm{k}\uparrow}
   -e^{i\theta_{\hat{\bm{k}}}}
   c^\dagger_{\bm{k}\downarrow}c^\dagger_{-\bm{k}\downarrow}
  \right)
 \end{array}
 \right.
\end{align}
The $\bm{d}$-vector for $A_\tx{tri}$ is
\begin{align}
 \bm{d}(\bm{k})
 &=-\frac12\tx{tr}\,
 \left[
  i\left(
 \begin{array}{cc}
  \frac{1}{2}e^{-i\theta_{\hat{k}}} & 0 \\
  0 & -\frac{1}{2}e^{i\theta_{\hat{k}}}
 \end{array}
 \right)\sigma_y\bm{\sigma}\right]\non
 &=-\frac{1}{2}\left(\cos{\theta_{\hat{k}}}\bm{e}_x
  +\sin{\theta_{\hat{k}}}\bm{e}_y\right)\non
 &=\frac{1}{2}\hat{\bm{k}}\times\bm{e}_z.
\end{align}
where we used Eq.~\eqref{eq:thetak} in the last line.

\begin{figure}[t!]
\includegraphics[width=\columnwidth]{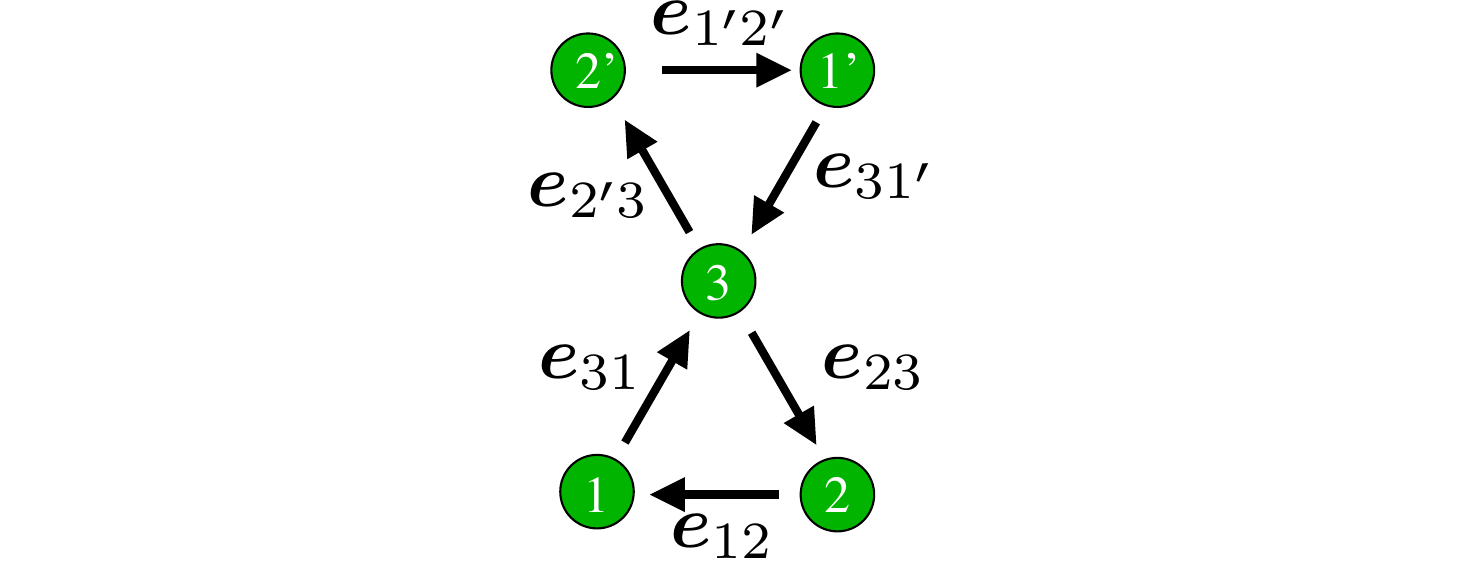}
 \caption{
 Schematic illustration of the unit vectors in Eq.~\eqref{eq:eXX}.
 }
 \label{fig:eXX}
\end{figure}
\section{Matrix elements of the BdG Hamiltonian}
\label{sec:BdGelements}
Let us rewrite the effective Hamiltonian in Eq.~\eqref{eq:effHam}:
\begin{align*}
 &H
 =-t\sum_{\langle ij\rangle}
 c_i^\dagger\bm{\chi}_i^\dagger
 \bm{\chi}_j c_j
 -(\mu+J_HS)\sum_{i}c_i^\dagger c_i\non
 &
 +\frac{1}{2}\sum_{i}\left(
 \tilde{\Delta}_s
 c_i^\dagger\bm{\chi}_i^\dagger
 i\sigma_y\bm{\chi}_i^*c_i^\dagger
 +\tx{h.c.}
 \right)
 \non
 &+\frac{1}{2}\sum_{\langle ij\rangle}
 \left(
 c_i^\dagger\bm{\chi}_i^\dagger
 \left(\Delta_s
 -i\Delta_p\left(\bm{e}_{ij}\times\bm{e}_z\right)\cdot\bm{\sigma}\right)
 i\sigma_y\bm{\chi}_j^*c_j^\dagger+\tx{h.c.}
 \right).
\end{align*}
Here, the third term disappears due to $c_i^\dagger c_i^\dagger=0$.
Now, we denote the creation operator by $c^\dagger_X(\bm{R})$, where $X$ is 
the label of sublattices and $\bm{R}$ is the position of the unit cell. By 
performing
Fourier transformation, i.e., defining
\begin{align}
 c^\dagger_X(\bm{k})
 =\frac{1}{\sqrt{S}}\sum_{\bm{R}}e^{-i\bm{k}\cdot\bm{R}}c^\dagger_X(\bm{R}),
\end{align}
where $S$ is the system size, the Hamiltonian reduces to Eq.~\eqref{eq:Heff},
which we rewrite here for convenience:
\begin{equation*}
 \begin{split}
  H
  &=\frac12\sum_{\bm{k}}\left(
  \bm{c}^\dagger(\bm{k})
  h(\bm{k})
  \bm{c}(\bm{k})+\tx{Tr\,}h_\tx{hop}(\bm{k})\right),\\
  h(\bm{k})
  &=\left(
  \begin{array}{cc}
   h_\tx{hop}(\bm{k}) & \Delta(\bm{k})\\
   \Delta^\dagger(\bm{k}) & -h_\tx{hop}^*(-\bm{k})
  \end{array}
  \right),
 \end{split}
\end{equation*}
where $\bm{c}^\dagger(\bm{k})
=\left(c^\dagger_1(\bm{k}),c^\dagger_2(\bm{k}),c^\dagger_3(\bm{k}),
c_1(-\bm{k}),c_2(-\bm{k}),c_3(-\bm{k})\right)$. Here,
the matrix elements of $h_\tx{hop}(\bm{k})$ and $\Delta(\bm{k})$ are given by
\begin{widetext}
\begin{align}
  h_\tx{hop}(\bm{k})
 &=-t\left(
 \begin{array}{ccc}
  0 & \alpha_{12}(1+e^{ik_1-ik_2}) & \alpha_{13}(1+e^{ik_1}) \\
  \alpha_{21}(1+e^{-ik_1+ik_2}) & 0 & \alpha_{23}(1+e^{ik_2}) \\
  \alpha_{31}(1+e^{ik_1}) & \alpha_{32}(1+e^{-ik_2}) & 0 
 \end{array}
 \right)-\mu\\
 \Delta(\bm{k})
 &=\left(
 \begin{array}{ccc}
  0 & \beta_{12}+\beta_{1'2'}e^{ik_1-ik_2}
   & \beta_{13}+\beta_{1'3}e^{ik_1}\\
  \beta_{21}+\beta_{2'1'}e^{-ik_1+ik_2} & 0 
   & \beta_{23}+\beta_{2'3}e^{ik_2}\\
  \beta_{31}+\beta_{31'}e^{-ik_1} 
   & \beta_{32}+\beta_{32'}e^{-ik_2} & 0
 \end{array}
 \right),
 \label{eq:deltak}
\end{align}
\end{widetext}
where
\begin{align}
 \alpha_{IJ}&=\bm{\chi}^\dagger_{I}\bm{\chi}_{J},\\
 \beta_{IJ}&=\bm{\chi}^\dagger_{I}
 \left(\Delta_s-i\Delta_p(\bm{e}_{IJ}\times\bm{e}_z)\cdot\bm{\sigma}\right)
 i\sigma_y
 \bm{\chi}_{J}^*.
\end{align}
Here, $\bm{\chi}_{I}$ is the spinor at sublattice $I(=1,2,3)$. (Recall that the
localized spins are aligned uniformly within each sublattice.)
The subscripts $1'$ and $2'$, appearing in Eq.~\eqref{eq:deltak}, are the 
``shifted'' sublattices obtained translation with the primitive vectors,
as illustrated in Fig.~\ref{fig:eXX}, and we use the following unit vectors:
\begin{equation}
 \begin{split}
  &\bm{e}_{12}=-\bm{e}_{21}=(-1,0)\\
  &\bm{e}_{23}=-\bm{e}_{32}=(1/2,-\sqrt{3}/2)\\
  &\bm{e}_{31}=-\bm{e}_{13}=(1/2,\sqrt{3}/2)\\
  &\bm{e}_{1'2'}=-\bm{e}_{2'1'}=(1,0)\\
  &\bm{e}_{2'3}=-\bm{e}_{32'}=(-1/2,\sqrt{3}/2)\\
  &\bm{e}_{31'}=-\bm{e}_{1'3}=(-1/2,-\sqrt{3}/2).
 \end{split}
 \label{eq:eXX}
\end{equation}

\section{Basic properties of the BdG system}
\label{sec:basic}
Here, we describe basic properties of the BdG Hamiltonian in 
Eq.~\eqref{eq:Heff}. We also discuss the possibility of negative 
quasiparticle energy.

For each momentum $\bm{k}$, the BdG eigenvalue equation reads
\begin{align}
 h(\bm{k})\left(
 \begin{array}{c}
  \bm{u}_n(\bm{k}) \\
  \bm{v}_n(\bm{k})
 \end{array}
 \right)
 =E_n(\bm{k})\left(
 \begin{array}{c}
  \bm{u}_n(\bm{k}) \\
  \bm{v}_n(\bm{k})
 \end{array}
 \right),
\end{align}
where $\bm{u}_n$ and $\bm{v}_n$ are three-component vectors. Here,
we consider only the three largest eigenvalues satisfying
\begin{align}
 E_1(\bm{k})\leq E_2(\bm{k})\leq E_3(\bm{k}).
\end{align}
Because of the particle-hole symmetry,
\begin{align}
 \Xi^{-1}h(\bm{k})\Xi&=-h(-\bm{k}),
\end{align}
where $\Xi=\tau_xK$ with $\tau_x$ the Pauli matrix for the Nambu space and $K$ 
complex conjugation, we also have
\begin{align}
 h(\bm{k})
 \left(
 \begin{array}{c}
  \bm{v}_n^*(-\bm{k}) \\
  \bm{u}_n^*(-\bm{k})
 \end{array}
 \right)
 =-E_n(-\bm{k})\left(
 \begin{array}{c}
  \bm{v}_n^*(-\bm{k}) \\
  \bm{u}_n^*(-\bm{k})
 \end{array}
 \right).
\end{align}
This implies that the six eigenvalues of $h(\bm{k})$ consist of
$E_n(\bm{k})$ and $-E_n(-\bm{k})$ with $n=1,2,3$. If the system 
preserves inversion symmetry, one has
$-E_n(-\bm{k})=-E_n(\bm{k})$, which guarantees that $E_n(\bm{k})$'s with 
$n=1,2,3$ are all positive. However, since our system lacks inversion symmetry,
this guarantee no longer holds, and $E_n(\bm{k})$ can become negative.
Indeed, $E_n(\bm{k})$ are negative for some parameters in 
Figs.~\ref{fig:PD} and \ref{fig:Eg}(a). We note the angle $\theta_\tx{min}$ 
in Fig.~\ref{fig:Eg}(b) always occurs in regions where $E_n(\bm{k})$ are
positive.

Using the quasiparticle operators defined by
\begin{align}
 \alpha_n^\dagger(\bm{k})=\bm{c}^\dagger(\bm{k})\left(
 \begin{array}{c}
  \bm{u}_n(\bm{k}) \\
  \bm{v}_n(\bm{k})
 \end{array}
 \right),
\end{align}
the BdG Hamiltonian is diagonalized as
\begin{align}
 H
 =\sum_{\bm{k}}\sum_{n=1}^3
 E_n(\bm{k})\alpha_n^\dagger(\bm{k})\alpha_n(\bm{k})+E_g.
\end{align}
Here, $E_g$ is the quasiparticle vacuum energy:
\begin{align}
 E_g
 =\frac12\sum_{\bm{k}}\tx{Tr\,}h_\tx{hop}(\bm{k})
 -\frac12\sum_{\bm{k}}\sum_{n=1}^3E_n(\bm{k}).
 \label{eq:Eg}
\end{align}
The expectation value of the particle number in the quasiparticle vacuum state
$\ket{\Omega}$ is
\begin{align}
 \bra{\Omega}N\ket{\Omega}
 =\sum_{\bm{k}}\sum_{n=1}^3\left|\bm{v}_{n}(\bm{k})\right|^2.
 \label{eq:N}
\end{align}

\begin{figure}
\includegraphics[width=\columnwidth]{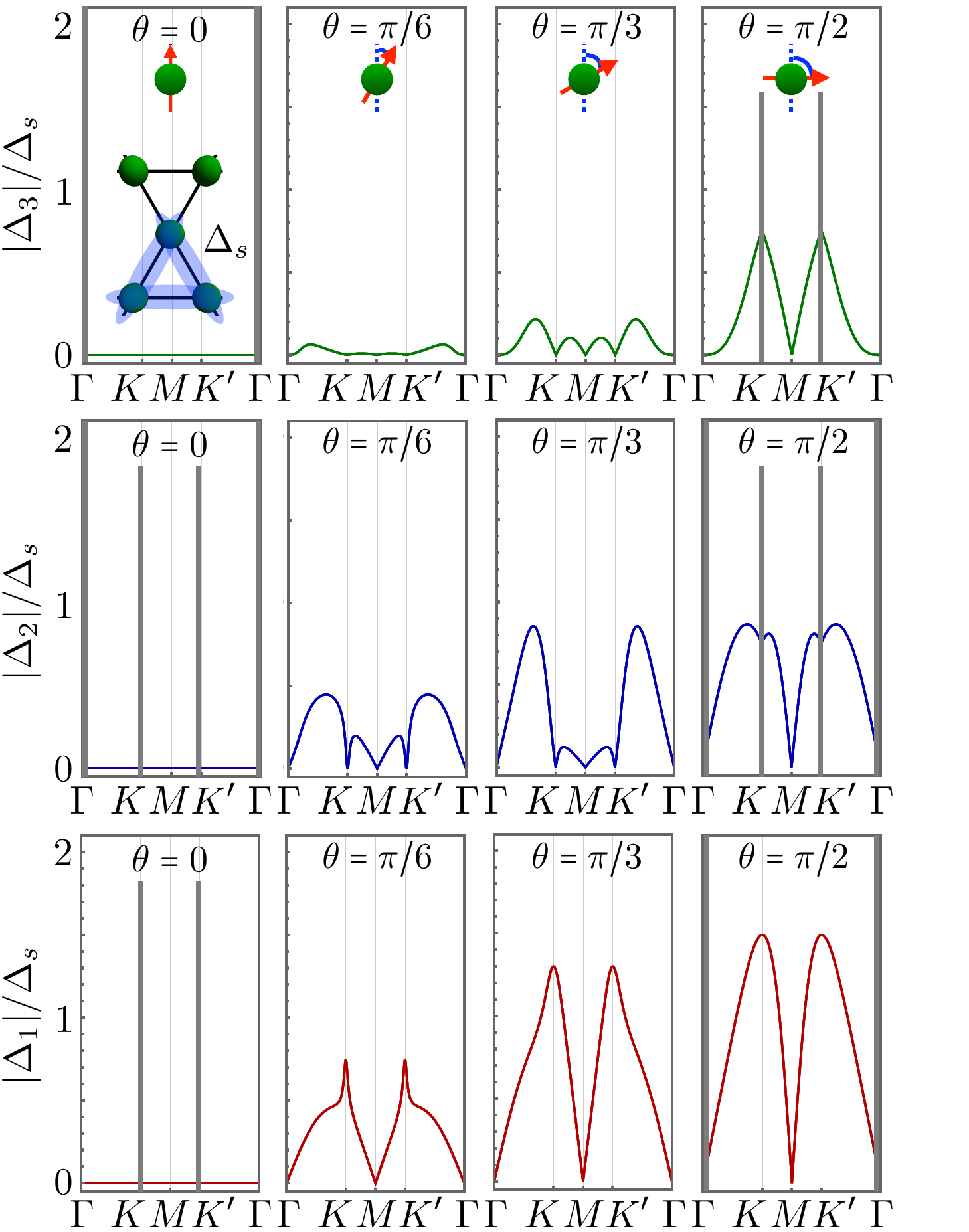}
 \caption{
 The same as Fig.~\ref{fig:pair-potential} but computed with
 Eq.~\eqref{eq:deltak-Chaou}. The inset in the top-left panel illustrates
 the NN $s$-wave pairing placed only on upward triangles.
 }
 \label{fig:pair-potential-Chaou}
\end{figure}
\section{Nonuniform $s$-wave pair potential}
\label{sec:Chaou-potential}
Here, we investigate the intraband $s$-wave pairing induced by breaking
inversion symmetry. To this end, we consider a system with  NN $s$-wave
pairing only on upward triangles of the kagome lattice~\cite{Chaou25},
see the inset of Fig.~\ref{fig:pair-potential-Chaou}. The pairing matrix in 
Eq.~\eqref{eq:deltak} is modified as
\begin{equation}
 \begin{split}
  \Delta(\bm{k})
 &=\left(
 \begin{array}{ccc}
  0 & \bar{\beta}_{12} & \bar{\beta}_{13} \\
  \bar{\beta}_{21} & 0 & \bar{\beta}_{23} \\
  \bar{\beta}_{31} & \bar{\beta}_{32} & 0
 \end{array}
  \right),\\
  \bar{\beta}_{IJ}
  &=i\Delta_s\bm{\chi}^\dagger_{I}
  \sigma_y\bm{\chi}_{J}^*.
  \label{eq:deltak-Chaou}
 \end{split}
\end{equation}
Note that $\bar{\beta}_{IJ}$ depends on $\phi_I-\phi_J$, and therefore is 
independent of $\phi$.
Figure~\ref{fig:pair-potential-Chaou} shows the pair potentials with
Eq.~\eqref{eq:deltak-Chaou} projected onto each band, where the finite 
intraband $s$-wave pairing is observed. 

\begin{figure}[t]
\includegraphics[width=\columnwidth]{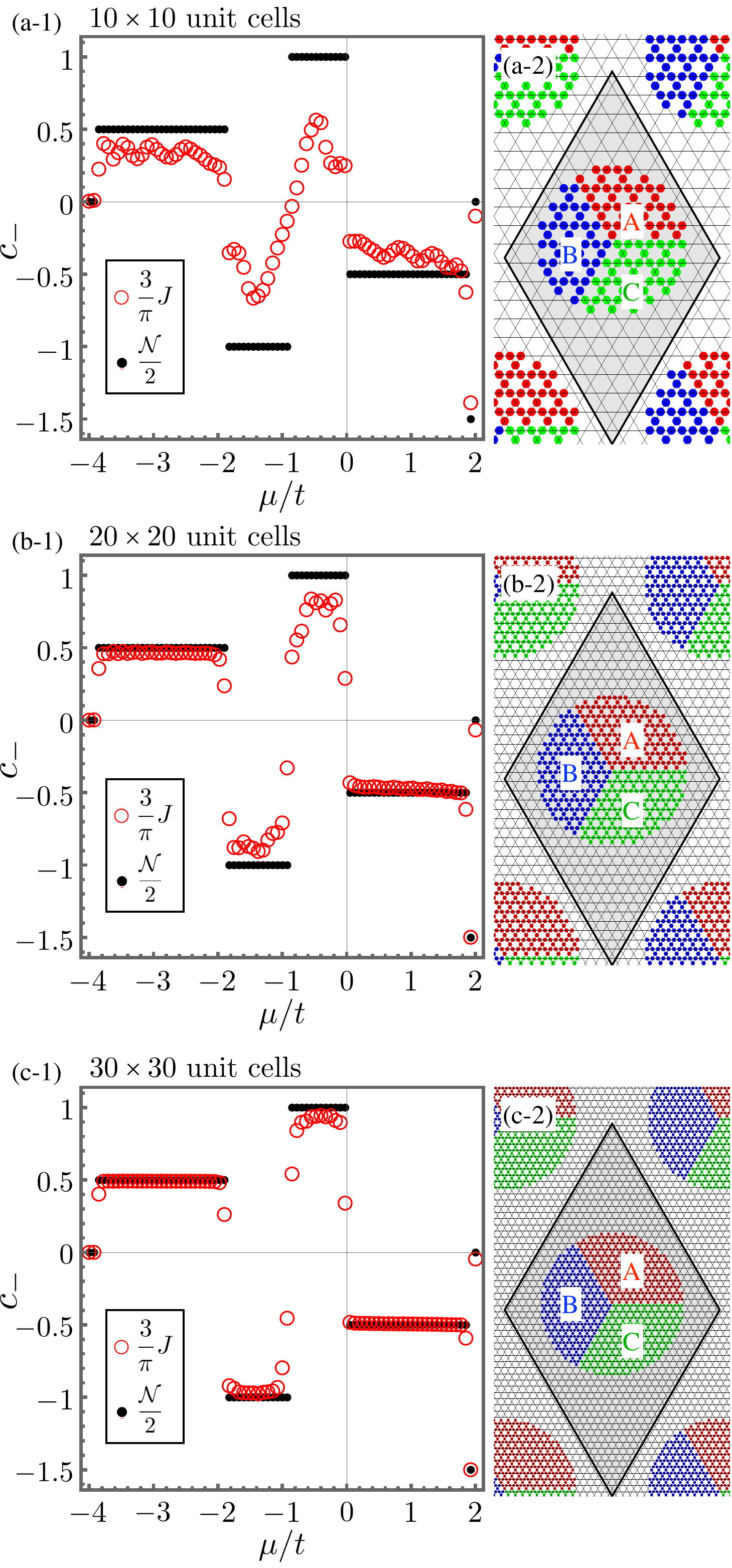}
 \caption{
 The same as Figs.~\ref{fig:ccc}(a) and (b) but for system sizes of 
 (a-1)(a-2) $10\times10$,
 (b-1)(b-2) $20\times20$ and (c-1)(c-2) $30\times30$ unit cells, respectively.}
 \label{fig:finite-size}
\end{figure}
\section{Finite-size effects on the modular commutator}
\label{appx:finite-size}
Here, we examine the dependence of the chiral central charge $c_-$ on the
system size. Figure~\ref{fig:finite-size} shows $c_-$ computed from the modular
commutator for system sizes of $10\times10$, $20\times20$, and $30\times30$ 
unit cells, along the same parameter line as Fig.~\ref{fig:ccc}(a). For the
smallest system $(10\times10)$, the quantization of $c_-$ is not clearly 
resolved. As the system size increases, the values converge toward the 
quantized plateaus, and the $40\times40$ lattice is sufficient for reliable
evaluation of $c_-$.

\bibliography{biblio_fqhe}

\begin{thebibliography}{147}
\bibitem [{Del()}]{Deltasp}%
  \BibitemOpen
  \href@noop {} {}\bibinfo {note} {The overall phase of the pairing term is a
  gauge freedom, allowing us to fix $\Delta_s$ to be a positive real number.
  Time-reversal symmetry constrains the relative phase between $\Delta_s$ and
  $\Delta_p$ to be either 1 or -1, which in our case corresponds to
  $\Delta_p=\pm|\Delta_p|$. The two choices are equivalent because the
  Hamiltonian with $(-\Delta_p,\bm{k})$ is identical to that with
  $(\Delta_p,-\bm{k})$: $\bm{d}(-\bm{k})=-\bm{d}(\bm{k})$ and all remaining
  terms are invariant under $\bm{k}\rightarrow-\bm{k}$.}\BibitemShut {Stop}%
\bibitem [{the()}]{thetaphi}%
  \BibitemOpen
  \href@noop {} {}\bibinfo {note} {The Hamiltonians with $(\theta,\phi)$ and
  $(\pi-\theta,\phi)$ are related by the reflection with respect to the $z=0$
  plane. The Hamiltonian with $(\theta,\phi)$ and $(\theta,\phi+\pi)$ are
  related by a $\pi$-rotation around a lattice site (noting that this rotation
  also changes the sign of the momentum.) Both transformations flip the sign of
  $\Delta_p$, but this can be compensated by the transformation
  $\bm{k}\rightarrow-\bm{k}$, as mentioned in Ref.~\cite{Deltasp}.}\BibitemShut
  {Stop}%
\end{thebibliography}

\end{document}